\documentclass[fleqn,usenatbib]{mnras}

\usepackage{newtxtext,newtxmath}

\usepackage[T1]{fontenc}

\DeclareRobustCommand{\VAN}[3]{#2}
\let\VANthebibliography\thebibliography
\def\thebibliography{\DeclareRobustCommand{\VAN}[3]{##3}\VANthebibliography}

\usepackage{graphicx}	
\usepackage{amsmath}	
\usepackage[version=4]{mhchem}
\usepackage{siunitx}
\usepackage{physics2}
\usephysicsmodule{ab}
\usepackage{derivative}

\newcommand{\Msun}{\mathrm{M}_\odot}
\newcommand{\Rsun}{\mathrm{R}_\odot}
\newcommand{\ej}{\mathrm{ej}}
\newcommand{\Ni}{\mathrm{Ni}}
\newcommand{\CSM}{\mathrm{CSM}}
\newcommand{\radni}{\ce{^{56}Ni}}

\title[Modelling Double-peaked SNe Ibc]{Hydrodynamic Modelling of Early Peaks in Type Ibc Supernovae with Circumstellar Interaction}

\author[R. Chiba et al.]{
Ryotaro Chiba,$^{1, 2}$\thanks{E-mail: ryotaro.chiba@grad.nao.ac.jp}
Takashi J. Moriya,$^{1, 2, 3}$
\\
$^{1}$Astronomical Science Program, Graduate Institute for Advanced Studies, SOKENDAI, 2-21-1 Osawa, Mitaka, Tokyo 181-8588, Japan\\
$^{2}$National Astronomical Observatory of Japan, National Institutes of Natural Sciences, 2-21-1 Osawa, Mitaka, Tokyo 181-8588, Japan\\
$^{3}$School of Physics and Astronomy, Monash University, Clayton, VIC 3800, Australia
}

\date{Accepted XXX. Received YYY; in original form ZZZ}

\pubyear{\the\year{}}

\begin{document}
\label{firstpage}
\pagerange{\pageref{firstpage}--\pageref{lastpage}}
\maketitle

\begin{abstract}
Recent high-cadence transient surveys have uncovered a subclass of Type Ibc supernovae (SNe) that exhibit an early, blue peak lasting a few days before the main, radioactively powered peak.
Since progenitors of Type Ibc SNe are typically compact and lack an extended envelope, this early peak is commonly attributed to the presence of circumstellar matter (CSM) surrounding the progenitor star.
As such, these SNe provide a unique opportunity to constrain the pre-explosion activity of Type Ibc SN progenitors.
We present the first systematic study of this Type Ibc SN population that incorporates hydrodynamic modelling.
We simulated Type Ibc SNe exploding within CSM using the multi-group radiation-hydrodynamics code \texttt{STELLA}, exploring a range of SN and CSM properties.
By comparing the theoretical multi-band light curves to a sample of seven Type Ibc SNe with early peaks, we constrained their CSM properties.
Assuming a wind-like density distribution of CSM, we found CSM masses of $10^{-2} - 10^{-1} \ \Msun$ and CSM radii of $(1 - 5) \times 10^3 \ \Rsun$.
While the masses were roughly consistent with a previous estimate obtained using an analytical model, the radii were significantly different, likely due to a simplified assumption on blackbody temperature used in analytical models. 
We infer that the progenitors could have created CSM via late-time binary mass transfer or pulsational pair instability.
We also estimate that, in the planned \textit{ULTRASAT} high-cadence survey, $\sim 30$ early peaks similar to those in this paper from Type Ibc SNe will be observed.
\end{abstract}




\section{Introduction}
\label{sec:intro}

It has been now well-established that the progenitors of many core-collapse SNe undergo enhanced mass loss prior to their explosions and explode within circumstellar matter (CSM) surrounding them. 
Evidence for this includes narrow spectral lines \citep[for recent reviews, see e.g.,][]{Fraser2020-xt, Dessart2024-rc}, pre-SN outbursts \citep[e.g.,][]{Pastorello2007-wn, Fraser2013-rk, Ofek2013-dr, Ho2019-nx, Strotjohann2021-qv, Jacobson-Galan2022-db}, and flash-ionisation features \citep[e.g.,][]{Gal-Yam2014-ez, Khazov2016-ld, Yaron2017-kn, Bruch2023-ks}.
However, the origin of the enhanced mass loss remains uncertain. 
Proposed mechanisms to explain this enhanced mass loss include gravity waves excited by core convection during the final burning stages \citep[e.g.,][]{Quataert2012-ex, Fuller2017-nv, Fuller2018-ya, Wu2021-kx}, pulsational pair instability \citep[e.g.,][]{Woosley2007-ti, Yoshida2016-mw, Woosley2017-bo}, and interaction with a binary companion \citep[e.g.,][]{Ouchi2017-ri, Wu2022-ek, Matsuoka2024-ho, Ercolino2024-bu, Ercolino2025-zx}.

Recent observations of some Type Ibc SNe whose light curves exhibit an early peak with a timescale of $\lesssim 10$ days that precedes the main, radioactively powered peak (e.g., SN 2008D, \citealt{Mazzali2008-xz}; LSQ 14efd, \citealt{Barbarino2017-yf}; iPTF 16hgs, \citealt{De2018-kv}; SN 2019dge, \citealt{Yao2020-qt}) may also indicate the presence of CSM around their progenitors \citep{Das2024-pm}.
These SNe resemble objects in a somewhat common subclass of Type IIb SNe with similar early peaks \citep[$\sim 30 \%$ among Type IIb SNe; ][]{Ayala2025-uy}, which are understood to be shock cooling emission from low-mass, extended hydrogen-rich envelopes of their progenitors \citep[e.g.,][]{Hoflich1993-gg, Shigeyama1994-hi, Woosley1994-ge, Blinnikov1998-kg, Arcavi2011-fl, Bersten2012-xm, Arcavi2017-wv}. 
However, given that the progenitors of Type Ibc SNe are understood to be compact Wolf--Rayet stars whose hydrogen-rich envelopes have been fully stripped, attributing early peaks in Type Ibc SNe to shock cooling emission would likely require the presence of CSM, which was then heated by the shock passing through it.
Shock cooling peaks have been shown to probe the mass and radius of the heated material \citep[e.g.,][]{Rabinak2011-uz, Nakar2014-yn, Piro2015-nm, Sapir2017-pw, Piro2021-cu, Margalit2022-mm, Morag2023-hm}, thus serving as a valuable tool for studying the environment surrounding SN progenitors.
Furthermore, the presence of CSM gives rise to extended shock breakout and emission from continued interaction, both of which can also contribute to the luminosity of early peaks \citep[e.g.,][]{Chevalier2011-xw, Khatami2024-ex, Nagy2025-jf}.

Other scenarios that may account for the early peaks include the expansion of He envelope during late-stage nuclear burning \citep[e.g.,][]{Kleiser2018-sl, Woosley2019-ke, Laplace2020-il, Wu2022-ek}, the presence of a $\radni$-rich shell in the outer layers of the progenitor \citep{Folatelli2006-ar, Bersten2013-vg, Orellana2022-xo}, and the weakening of gamma-ray burst jets due to CSM \citep{Hamidani2025-bq}. 

Recently, \citet{Das2024-pm} conducted a systematic analysis of this population of Type Ibc SNe using SN samples obtained by the Zwicky Transient Facility (ZTF). 
Their work yielded the identification of 17 such SNe, and the estimate of the fraction of Type Ibc SNe with an early peak was determined to be $3 - 9 \%$. 
The authors also obtained estimates of the CSM masses and radii that account for the observed early peaks using order-of-magnitude estimates and light curve fitting to the analytical model of shock cooling emission from extended envelopes by \citet{Piro2021-cu}.
In this study, we further investigate this population of Type Ibc SNe with early peaks, utilising a multi-group radiation-hydrodynamic modelling code by assuming that the early peaks are powered by interaction with CSM.
Through the consistent fitting of both peaks in the light curves, we estimate the physical parameters of the explosion and CSM, which we then use to discuss potential mechanisms for CSM formation.

The selection of the sample and the details of the numerical simulation are described in Section~\ref{sec:methods}. 
Results of our simulation and the inferred CSM configuration are presented in Section~\ref{sec:result}. 
The mass loss scenarios and the potential for future detection of early peaks are discussed in Section~\ref{sec:discussion}. 
Finally, we summarise our findings in Section~\ref{sec:conclusion}.

\section{Methods}
\label{sec:methods}

\subsection{Data selection}
\label{sec:data}

From the 17 Type Ibc SNe with early peaks collected by \citet{Das2024-pm}, we select a subsample of seven SNe for which the explosion energy ($E_\ej$) is constrained by the photospheric velocity ($v_\mathrm{ph}$) around the radioactively powered main peak. 
This ensures that the degeneracy between explosion energy, ejecta mass ($M_\ej$) and $\radni$ mass ($M_\mathrm{Ni}$) \citep{Arnett1982-lb, Wheeler2015-nl} is resolved, allowing for a consistent fit of both the CSM interaction powered and radioactively powered peaks. 
As previously highlighted by \citet{Piro2015-nm}, a degeneracy between the CSM radius and SN explosion energy exists in the behaviour of shock cooling light curves. 
Consequently, a robust determination of the explosion energy is crucial for adequately constraining the CSM properties responsible for the early peaks. 
We excluded SN 2021inl from the sample, as its early peak was significantly brighter than the radioactively powered main peak, making it difficult to estimate the contribution of radioactive heating alone.
The details of the selected SNe are listed in Table~\ref{tab:sample_info}.

\subsection{Details of numerical modelling and initial conditions}
\label{sec:init_cond}

In order to properly model the hydrodynamic and radiative processes involved in the interaction between SN ejecta and CSM, the one-dimensional multi-group radiation-hydrodynamics code \texttt{STELLA} \citep{Blinnikov1998-kg, Blinnikov2000-xz, Blinnikov2006-hz, Blinnikov2011-wy} is employed. 
\texttt{STELLA} has previously been used to model the light curves of Type Ibc SNe with CSM interaction \citep[e.g.,][]{Baklanov2015-fg, Jin2021-hq}. 
\texttt{STELLA} calculates the evolution of spectral energy distributions (SEDs) at each time step and for each Lagrangian mass mesh by solving radiative transfer with the variable Eddington method. 
The multi-colour light curves are obtained by convolving the obtained SEDs with filter functions. 

For each calculation, Lagrangian mass meshes are taken fine enough to ensure the convergence.
We used the standard 100 frequency bins in the SED calculations, which range from $1 - 50,000$ Å on a logarithmic scale.
When calculating the opacity table, various processes including bremsstrahlung, bound–free transitions, bound–bound transitions, photoionization, and electron scattering are treated within the frequency range \citep[for details, see e.g.,][]{Kozyreva2020-do}.

We note that our model does not fully describe the effects of bremsstrahlung and inverse Compton scattering when the collisionless shock that emerges after the shock breakout heats the post-shock plasma to the temperature of $\sim \qty{100}{keV}$.
However, Compton degradation and bound-free absorption within CSM severely suppress the X-ray emission until the shock reaches the outer edge, thermalising most of the X-ray emission to optical wavelengths \citep{Chevalier2012-ir, Dessart2015-uj, Margalit2022-gx}.
Although complete photoionisation of CSM prior to the shock reaching the CSM outer edge may allow the X-ray photons to escape, such a state is achieved only for either relativistic shock velocities or sufficiently low-density CSM, neither of which is satisfied in our relevant models \citep{Margalit2022-gx}.
We therefore consider that our models suffice in calculating optical light curves even when the shock is collisionless and photons with energies beyond the wavelength range that we have considered are produced, although we note there may be increased uncertainty in cases where the CSM density approaches the low-density regime.

The initial conditions of the models are based on the hydrodynamic structure of CO21 model \citep{Iwamoto1994-lb}, which is a canonical hydrodynamic profile of a stripped-envelope explosion with $M_\ej = 0.86 \ \Msun$, $M_\Ni = 0.081 \ \Msun$, and $E_\ej = \qty{0.91e51}{erg}$. 
Although this model is based on the explosion of a C+O star, since envelopes of Type Ibc SN progenitors are both radiative and have similar density profiles, this model should be applicable for modelling of Type Ib SNe as well.
Assuming homologous expansion of the ejecta ($v = r / t$), we construct the density profile when the outermost radius of the expanding ejecta is $\qty{5e12}{cm}$, corresponding to $\sim \qty{0.02}{d}$ after the explosion.
While this assumption neglects the earliest portion of the interaction, it is justified because $t \sim \qty{0.02}{d}$ is much shorter than the typical rise time of early peaks.

\begin{figure}
    \centering
    \includegraphics[width=0.9\linewidth]{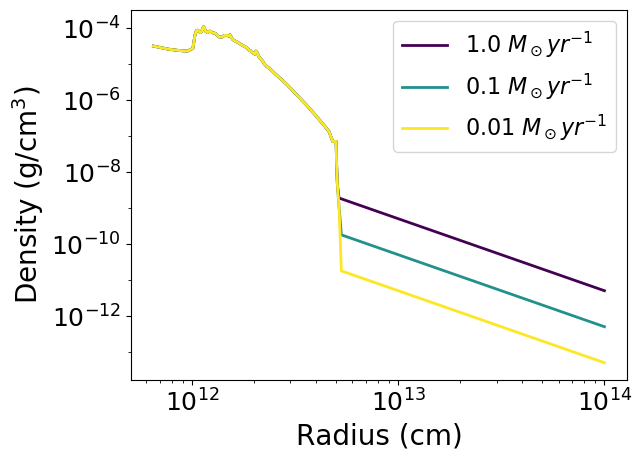}
    \caption{Examples of initial density profiles used in our modelling. 
    Wind-like CSM ($\rho_\CSM \propto r^{-2}$) with a fixed $v_\mathrm{w} = \qty{1000}{km.s^{-1}}$ and radius $R_\CSM = 10^{14} \ \mathrm{cm}$ and different values of $\dot{M}$ is attached on top of ejecta with mass $M_\mathrm{ej} = 1 \ \Msun$ at $\qty{0.02}{d}$ after the explosion.} 
    \label{fig:density_profiles}
\end{figure}

Density profiles corresponding to a different set of parameters of explosion (${M_\ej}^\prime$ and ${E_\ej}^\prime$) are then obtained by scaling the density $\rho$ and velocity $v$ at each mass coordinate, as follows:
\begin{align}
    \rho^\prime &= \rho \ab (\frac{{M_\ej}^\prime}{M_\ej}), \label{eq:new_density_profile} \\
    v^\prime &= v \ab (\frac{{M_\ej}^\prime}{M_\ej})^{-1/2} \ab (\frac{{E_\ej}^\prime}{E_\ej})^{1/2}. \label{eq:new_velocity_profile}
\end{align}
This assumption is motivated by the fact that the expanding SN ejecta is known to approximately follow a double power-law density profile with similar power indices \citep{Matzner1999-lt}.
Only the interaction between the CSM and the outer ejecta, which is well described by single power-law density profile (as can also be noted in Figure~\ref{fig:density_profiles}), is responsible for early peaks we consider here.

In order to justify the simplifying assumptions in setting the initial condition, we ran a test calculation and compared the results with those of \citet{Jin2021-hq}, who used the thermal bomb method to explode the progenitors obtained via stellar evolution simulation. 
The results are shown in Appendix~\ref{sec:Jin21_test}; we found that the results were sufficiently consistent for the purpose of this work.

The amount of $\radni$ is adjusted through modification of the abundance profile of the model. 
A double-peaked $\radni$ distribution has also been proposed as a mechanism to explain early peaks in SNe. However, since the focus of this study is on the contribution of CSM interaction alone, either of the following simple treatments is assumed.

\vspace{-2mm}
\begin{itemize}
    \item ``Box-shaped'' distribution: the region with $X_\mathrm{Ni} = 0.75$ extends from the fixed mass cut $M_\mathrm{cut}$ (corresponding to the iron core mass) up to a mass coordinate $M_\mathrm{top}$ selected so that $X_\mathrm{Ni} (M_\mathrm{top} - M_\mathrm{cut}) = M_\Ni$ for a desired $M_\Ni$. Outwards of $M_\mathrm{top}$, $X_\mathrm{Ni}$ is set to 0.

    \item ``Uniform'' distribution: $\radni$ is evenly distributed throughout the ejecta from $M_\mathrm{cut}$ up to $M_\mathrm{cut} + M_\ej$, so that $X_\mathrm{Ni} M_\ej = M_\Ni$ for a desired $M_\Ni$.
\end{itemize}
\vspace{-2mm}
In this work, the amount and distribution of $\radni$ does not significantly affect the appearance of early peaks, since radioactive heating occurs on a significantly slower timescale than that of the early peaks.

It is assumed that the CSM follows a wind-like density profile:
\begin{equation}
    \rho_\CSM = \frac{\dot{M}}{4 \pi v_\mathrm{w} r^2},
\end{equation}
where $\dot{M}$ is the mass-loss rate, $v_\mathrm{w}$ is the wind velocity and $r$ is the radius. 
Here, $v_\mathrm{w}$ is fixed at a constant value of $\qty{1000}{km.s^{-1}}$, which corresponds to the typical wind velocity of Wolf--Rayet stars. 
CSM extends from the outer edge of the SN ejecta at $\qty{5e12}{cm}$ to a given outer radius $R_\CSM$. 
Figure~\ref{fig:density_profiles} shows the initial CSM density structure in some of our models.

CSM is assumed to have the same composition as the outermost layer of the original CO21 model, with $X_\mathrm{C} = 0.45$, $X_\mathrm{O} = 0.44$. 
It is important to note that, given the primary contribution to opacity within hydrogen-free CSM is electron scattering, opacity is largely independent of composition, as long as it is hydrogen-free.

\begin{table*}
    \centering
    \begin{tabular}{ccc|cccc|ccc}
        \hline
        Event & Redshift & Type & $E_\ej$ & $M_\ej$ & $M_\Ni$ & $\radni$ & $\dot{M}$ & $M_\CSM$ & $R_\CSM$\\
         &  &  & ($10^{51} \ \mathrm{erg}$) & ($\Msun$) & ($\Msun$) & Dist & ($\Msun \ \mathrm{yr}^{-1}$) & ($\Msun$) & ($\Rsun$)\\
        \hline
        SN 2018lqo & 0.033 & Ib     & 0.61 & 1.0 (1.53) & 0.03 (0.031) & Box & 3 & 0.062 & 1000 \\
        SN 2020bvc & 0.025 & Ic-BL  & 6.70 & 2.5 (3.47) & 0.6 (0.34) & Box & 3 & 0.090 & 1400 \\
        SN 2020kzs & 0.037 & Ib     & 0.93 & 2.0 (3.04) & 0.2 (0.14) & Box & 3 & 0.19 & 2900 \\
        SN 2021gno & 0.006 & Ib     & 0.27 & 0.8 (0.71) & 0.009 (0.010) & Box & 0.5 & 0.010 & 1000 \\
        SN 2021nng & 0.040 & Ib     & 2.38 & 7.0 (10.35) & 0.4 (0.67) & Box & 1 & 0.094 & 4300 \\
        SN 2021aabp & 0.064 & Ic-BL & 2.25 & 2.0 (3.57) & 0.7 (0.62) & Box & 2 & 0.19 & 4300 \\
        SN 2022oqm & 0.011 & Ic     & 0.23 & 1.0 (0.85) & 0.09 (0.089) & Uni & 0.5 & 0.047 & 4300 \\
        \hline
    \end{tabular}
    \caption{Summary of the SN sample and its best fit models. 
    $E_\ej$ is the value estimated by \citet{Das2024-pm} by applying the relation outlined by \citet{Arnett1989-bk} between photospheric velocity at peak, rise time and explosion energy $E_\ej$. 
    Ejecta mass $M_\ej$, $\radni$ mass $M_\Ni$, total CSM mass $M_\CSM$, and CSM radius $R_\CSM$ are obtained via light curve fitting. 
    Values in parentheses are $M_\ej$ and $M_\Ni$ reported in \citet{Das2024-pm} using \citet{Arnett1989-bk} model.
    In the column ``$\radni$ Dist'', ``Box'' indicates a box-shaped, ``Uni'' indicates an uniform distribution of $\radni$.
    }
    \label{tab:sample_info}
\end{table*}

\subsection{Analysis and parameter space}
\label{sec:analysis}

For each SN in the sample with ejecta energy $E_\ej$ constrained by its spectra, we first estimate its $M_\ej$ and $M_\Ni$ by fitting its radioactively powered peak. 
This is because the values in \citet{Das2024-pm} are obtained by using analytic formulae in \citet{Arnett1989-bk}. 
Since they do not necessarily achieve the optimal fit to the observed light curve with the use of hydrodynamic modelling, the values of $M_\ej$ and $M_\Ni$ from the same study are not used here.
While $E_\ej$ is not fitted simultaneously with other parameters, we ensure that its choice is consistent with the choice of other parameters, as outlined later.
Then, using early peaks of the SNe, we infer the properties of CSM configuration ($\dot{M}$ and $R_\CSM$).
The details of the procedure are as follows.

Models of SN ejecta are constructed with the explosion energy $E_\ej$ fixed to the values estimated in \citet{Das2024-pm}.
We consider ejecta masses ranging from $M_\ej = 0.1 - 10 \ \Msun$ and $\radni$ masses ranging from $M_\Ni = 0.007 - 0.7 \ \Msun$.
Since our focus here is to ensure consistency between CSM interaction powered and radioactively powered peaks, we only tried to achieve a rough fit to the radioactively powered peak and did not make more precise adjustments of the initial conditions (e.g., tuning the $\radni$ profile, changing the ejecta density profile).
For instance, SN 2022oqm exhibited three distinct light curve peaks, of which \citet{Yadavalli2024-wg} attributed the first to shock cooling emission from CSM, and the remaining two to radioactive decay from two distinct sources of $\radni$ \citep[but see also][]{Irani2024-zl}.
Based on their interpretation, we only used the third, most pronounced peak to constrain the parameters of explosion, disregarding the less prominent contribution of the second peak.

Table~\ref{tab:sample_info} shows the values of $M_\ej$ and $M_\Ni$ that we obtained in this way, along with the values estimated by \citet{Das2024-pm} using the \citet{Arnett1989-bk} model. 
We see that both sets of the values are roughly consistent, with $M_\ej$ being in agreement within a factor of $1.5$.
This ensures that the value of $\sqrt{2 E_\ej / M_\ej}$ remains comparable to the value of $v_\mathrm{ph}$ at radioactively powered main peak
and validates our approach to set the values of $E_\ej$ to those reported in \citet{Das2024-pm}.

Once $M_\ej$ and $M_\Ni$ for each SN have been determined, we then construct models for the entire light curves by attaching CSM around the SN ejecta. 
We consider mass-loss rates ranging from $\dot{M} = 0.01 - 7 \ \Msun \ \mathrm{yr}^{-1}$ and CSM radii ranging from $R_\CSM = 10^{13} - 10^{15}~\mathrm{cm}$. 

\section{Results}
\label{sec:result}

\subsection{Relation between various parameters and the appearance of early peaks}
\label{sec:dependence}

In this section, we show how the early peaks in our model scale with the parameters of explosion ($M_\ej$, $E_\ej$) as well as the parameters of CSM configuration ($\dot{M}$, $R_\CSM$).

\begin{figure*}
    \centering
    \includegraphics[width=1\linewidth]{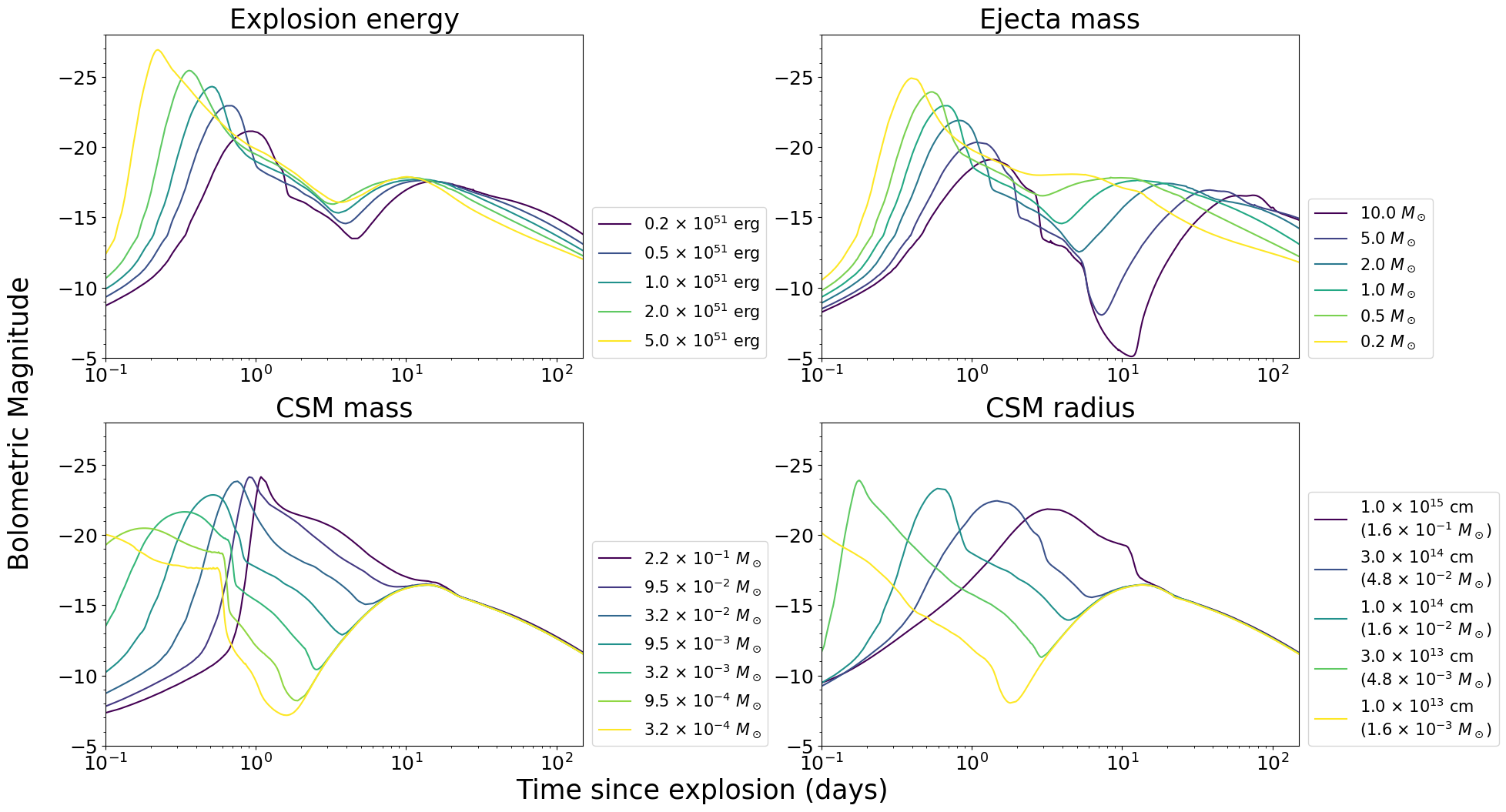}
    \caption{
    Upper panels: 
    Bolometric light curves obtained by varying the parameters of explosion ($M_\ej$, $E_\ej$) while keeping the configuration of CSM identical.
    The CSM configuration is fixed to $\dot{M} = 0.5 \ \Msun \ \mathrm{yr}^{-1}$ (corresponding to $M_\CSM = 0.016 \ M_\odot$) and $R_\CSM = \qty{1e14}{cm}$.
    In the left panel, explosion energy $E_\ej$ is varied while ejecta mass is set to $M_\ej = 1 \ \Msun$.
    In the right panel, ejecta mass $M_\ej$ is varied while the explosion energy is set to $E_\ej = 1 \times 10^{51} \ \mathrm{erg}$. \\
    Lower panels:
    Bolometric light curves obtained by varying the parameters of CSM configuration ($\dot{M}$, $R_\CSM$) while keeping the parameters of explosion identical.
    The parameters of explosion were set to those estimated for SN 2018lqo ($E_\ej = 0.61 \times 10^{51} \ \mathrm{erg}$, $M_\ej = 1.0 \ \Msun$, $M_\Ni = 0.03 \ \Msun$), while CSM radius is set to $R_\CSM = \qty{1e14}{cm}$.
    In the left panel, mass-loss rate $\dot{M}$ (and in turn total CSM mass $M_\CSM$) is varied while the CSM radius is set to $R_\CSM = 10^{14} \ \mathrm{cm}$.
    In the right panel, $R_\CSM$ is varied while the mass-loss rate is set to $\dot{M} = 0.5 \ \Msun  \ \mathrm{yr}^{-1}$.
    }
    \label{fig:Mbol_LC_all}
\end{figure*}


In the top row of Figure~\ref{fig:Mbol_LC_all}, we show bolometric light curves for the fixed configuration of CSM ($R_\CSM = \qty{1e14}{cm}$ and $\dot{M} = 0.5 \ \Msun  \ \mathrm{yr}^{-1}$, corresponding to the total CSM mass $M_\CSM = 0.016 \ M_\odot$) and different parameters of explosion.
In the bottom row of Figure~\ref{fig:Mbol_LC_all}, we similarly show a series of bolometric light curves for the fixed parameters of explosion ($E_\ej = 0.61 \times 10^{51} \ \mathrm{erg}$, $M_\ej = 1.0 \ \Msun$, $M_\Ni = 0.03 \ \Msun$ with ``box-shaped'' distribution) and different parameters of CSM configuration ($\dot{M}$, $R_\CSM$).

We can compare these light curves with the analytical calculation of \citet{Khatami2024-ex}.
They found that the morphology of bolometric light curves of interaction-powered transients is characterised by the mass ratio between CSM and ejecta, and by the location of shock breakout.
Since these models satisfy $M_\ej > M_\CSM$, they fall under either of the following two regimes in their nomenclature.

\begin{itemize}
    \item 
    \textbf{Light CSM, Interior Breakout}.
    CSM is optically thin enough so that shock breakout occurs far from the edge of CSM. 
    In this case, after the shock breakout, the light curve plateaus due to continued interaction until the shock reaches the edge of CSM.
    Thereafter, the luminosity drops and transitions to shock cooling phase.
    
    \item 
    \textbf{Light CSM, Edge Breakout}.
    The shock breakout occurs near the edge of CSM.
    In this case, after shock breakout, luminosity drops without exhibiting a plateau.
\end{itemize}

In the bottom left panel of Figure~\ref{fig:Mbol_LC_all}, we can see how increasing the CSM mass and therefore its optical depth leads to the transition from the ``internal breakout'' to the ``edge breakout'' regime.
For models with $M_\CSM < 10^{-3} \ M_\odot$, light curves are in the ``internal breakout'' regime.
The light curve exhibits a shock breakout peak at $t \sim 0.5 \ \mathrm{d}$ and then a plateau lasting until $t \sim 1 \ \mathrm{d}$ when the shock emerges from the edge of CSM.
For larger values of $M_\CSM$ light curves are closer to the ``edge breakout'' regime and the plateaus are less pronounced.
Note that shock cooling phases do not manifest themselves as light curve peaks in the bolometric light curves.

\begin{figure*}
    \centering
    \includegraphics[width=1\linewidth]{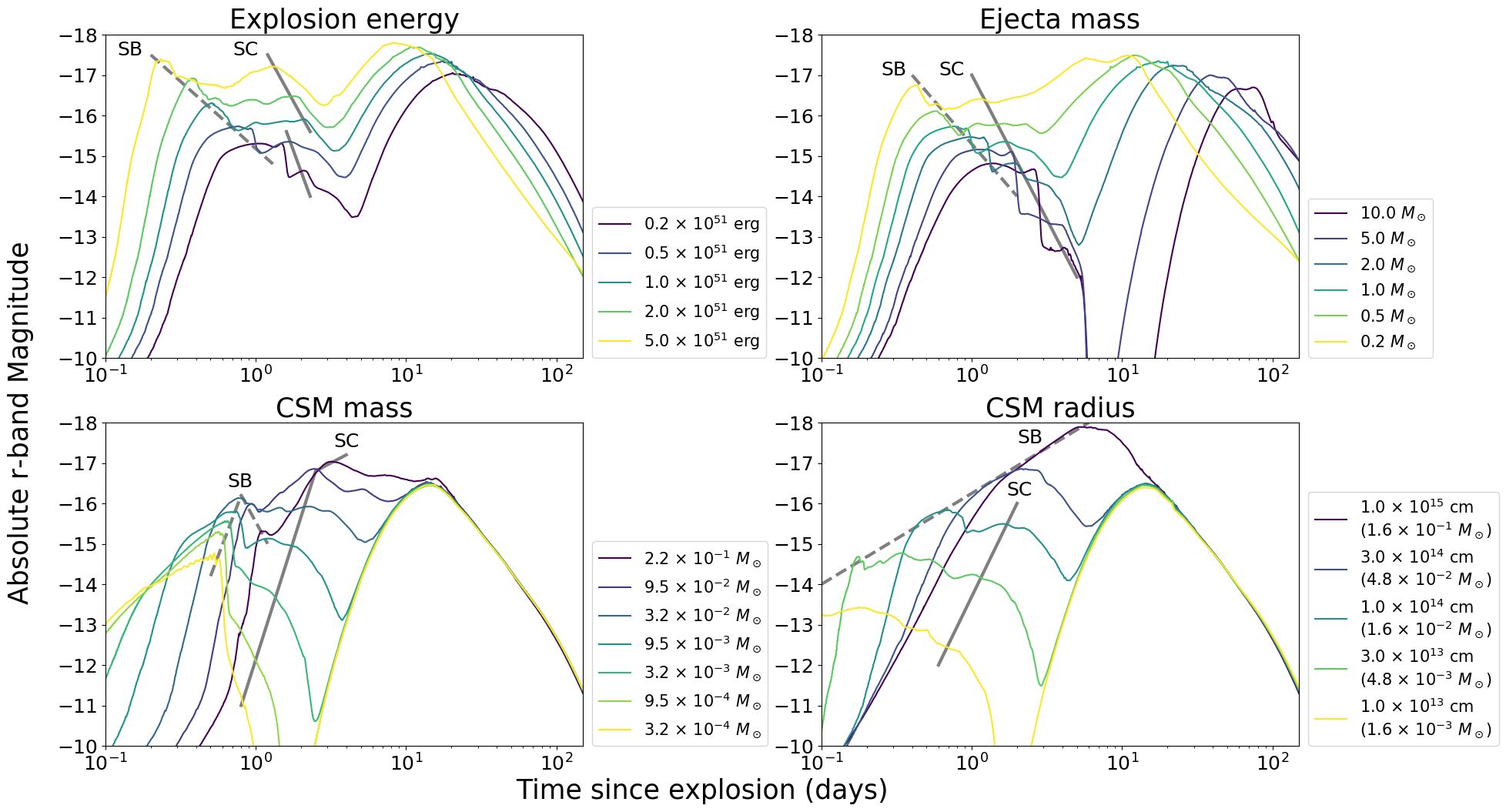}
    \caption{
    Same as Figure~\ref{fig:Mbol_LC_all}, but showing $r$-band light curves.
    In addition, approximate locations of shock breakout peaks (``SB'', dashed) and shock cooling peaks (``SC'', solid) within the light curves are highlighted.
    }
    \label{fig:Mr_LC_all}
\end{figure*}

Figure~\ref{fig:Mr_LC_all} shows a series of $r$-band light curves for the same set of models as Figure~\ref{fig:Mbol_LC_all}.
Unlike bolometric light curves, shock cooling phase in $r$-band (or optical bands in general) can manifest itself as an additional light curve peak that lie between the shock breakout peak and the radioactively powered peak.
This is because the temperature of the shock-heated material drops as it cools down and the dominant wavelength of its emission approaches the optical range.
In the top row of Figure~\ref{fig:Mr_LC_all}, we can see that the most of the light curves exhibit the shock breakout peaks ($t \sim 0.1 - 1 \ \mathrm{d}$) followed by peak/excess due to shock cooling emission ($t \sim 1-5 \ \mathrm{d}$).
The bottom left panel of Figure~\ref{fig:Mr_LC_all} shows that progressively larger values of $M_\CSM$ lead to greater relative prominence of shock cooling peaks compared to shock breakout peaks.
Figure~\ref{fig:Mr_LC_all} also shows how time and $r$-band magnitude of shock cooling peaks (marked with solid, grey lines) both scale with respect to different parameters.
If the parameters of explosion ($E_\ej$ and $M_\ej$) are well-constrained, time and magnitude of early peaks are the most significant parameters that constrain the properties of CSM configuration ($\dot{M}$ and $R_\CSM$).
Timescales of shock cooling peaks, which is determined from the balance between expansion and diffusion timescales of shocked material, is predominantly affected by $M_\CSM$ for the same parameters of explosion \citep{Arnett1982-lb}.
Then, for the same value of $M_\CSM$, the extent of adiabatic loss from the shocked material (and therefore $R_\CSM$) determines the peak luminosity.


\begin{figure}
    \centering
    \includegraphics[width=1\linewidth]{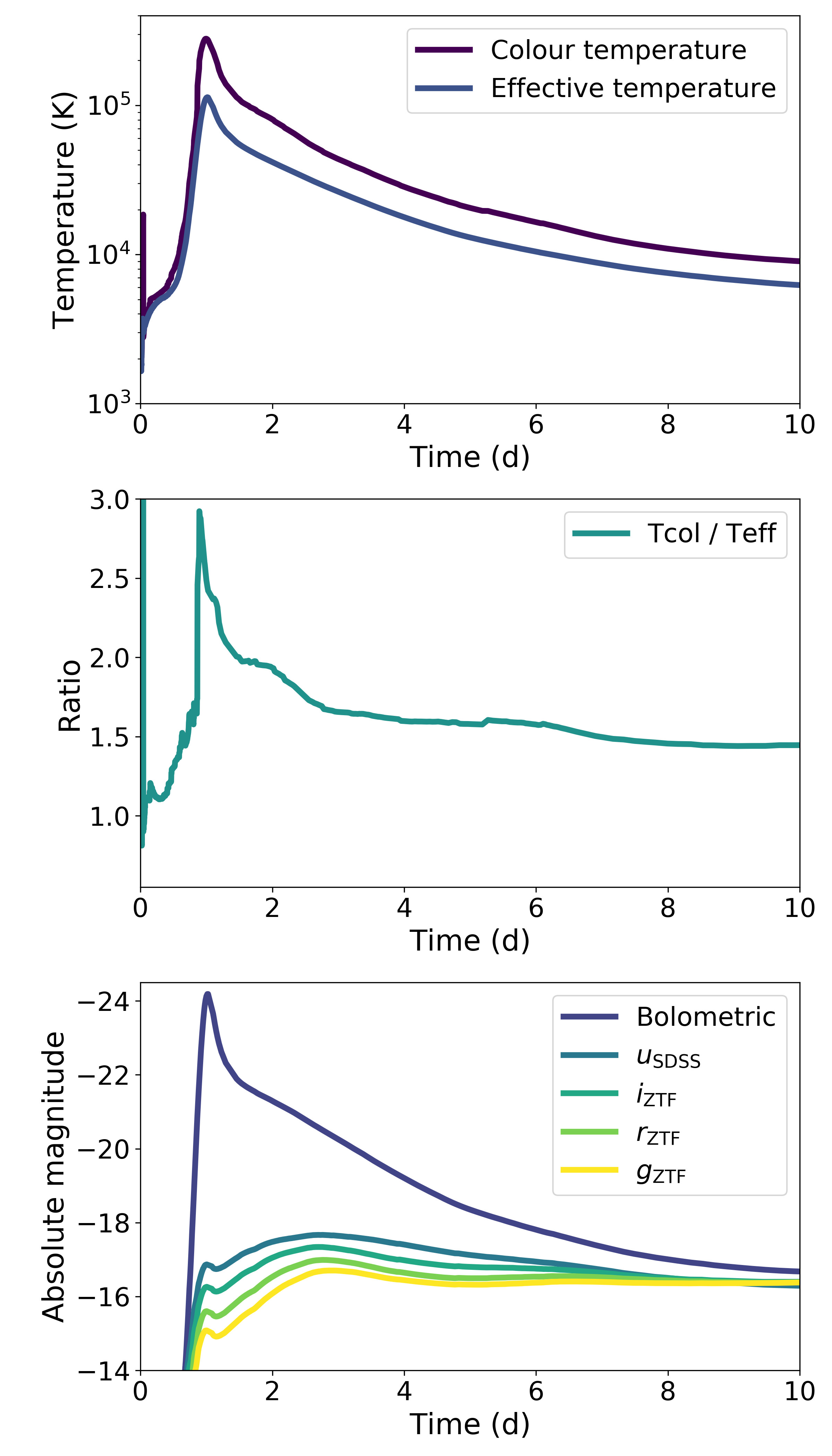}
    \caption{
    The evolution of the effective and colour temperatures, ratio between the two, and the bolometric and $r$-band light curves for the model with $E_\ej = 0.61 \times 10^{51} \ \mathrm{erg}$, $M_\ej = 1.0 \ \Msun$, $R_\CSM = \qty{1e14}{cm}$, $M_\CSM = 0.1 \ M_\odot$.
    }
    \label{fig:temperature}
\end{figure}

Figure~\ref{fig:temperature} shows the typical evolution of effective temperature and colour temperature of the shock cooling emission in our models.
Here, colour temperature is obtained by fitting the SED of the emitted photons at each moment with a blackbody.
Effective temperature is obtained as the gas temperature of the photosphere, or, the layer at which the optical depth calculated with Rosseland mean opacity, $\tau$, is equal to $2/3$.
We find the colour temperature during the shock cooling peak in $r$-band ($t = 2-4 \ \mathrm{d}$ in this model) to be $1.5-2$ times higher than the photospheric temperature.
This is in contrast with the assumption made in most analytic models of shock cooling emission that the shock cooling emission radiates as a blackbody whose temperature is equal to the photospheric temperature \citep[e.g.,][]{Piro2015-nm, Piro2021-cu, Margalit2022-gx}.
The figure shows that photons emitted during the shock cooling phase are supplied from inner, slightly hotter region of the shocked material.
The photons are then mostly scattered, rather than absorbed, through the shocked CSM that are hot and ionised, without changing their wavelength significantly.
Since the visual band is in the Rayleigh-Jeans regime, the visual flux in our models becomes fainter compared to the analytical models by an order of magnitude.

\subsection{Estimated parameters of CSM}
\label{sec:parameters}

\begin{figure*}
    \centering
    \includegraphics[width=1\linewidth]{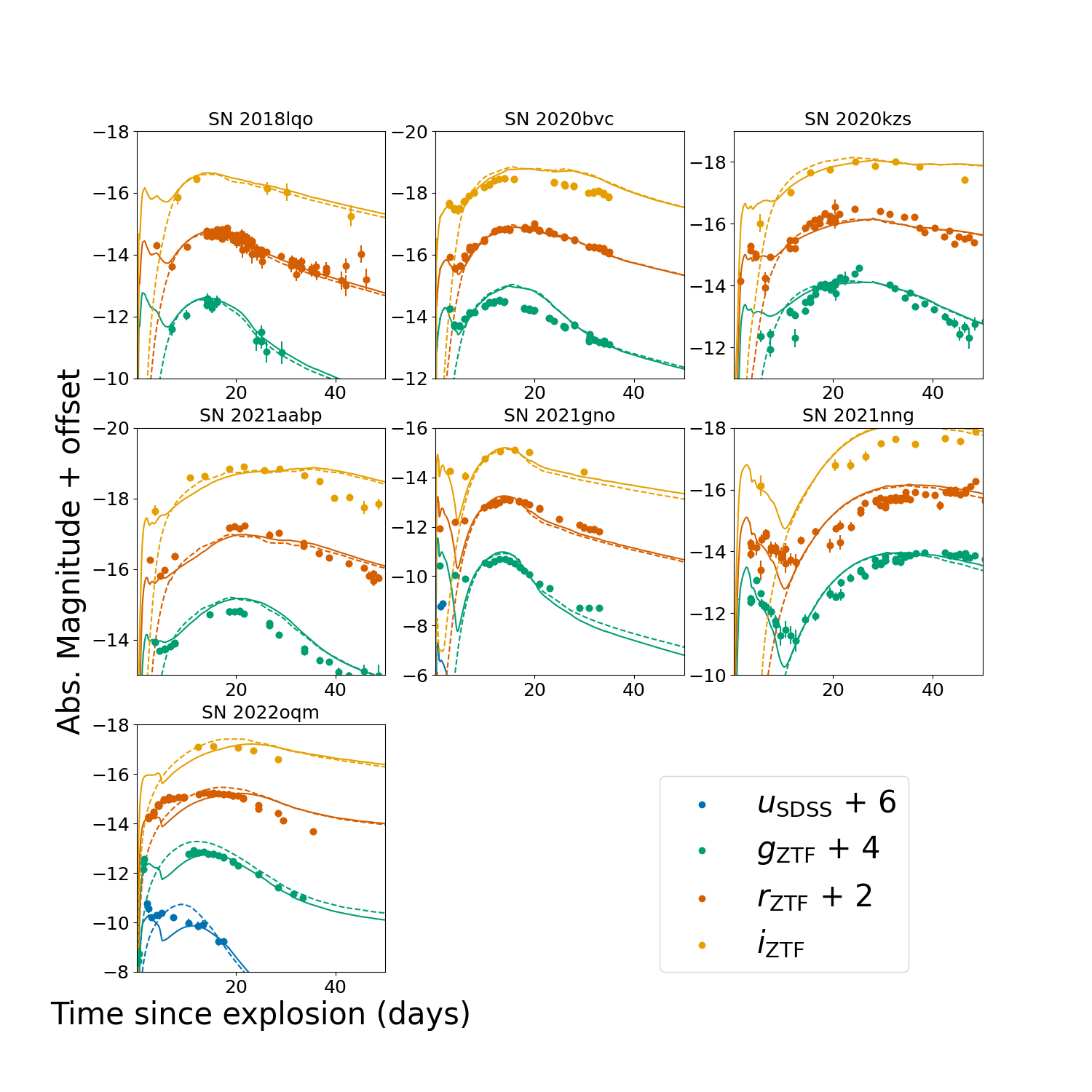}
    \caption{Synthetic light curves of the best fit models (solid lines), synthetic light curves of the best fit models without CSM (dotted lines), and the observed light curves (dots).}
    \label{fig:lc_all}
\end{figure*}

\begin{figure}
    \centering
    \includegraphics[width=1\linewidth]{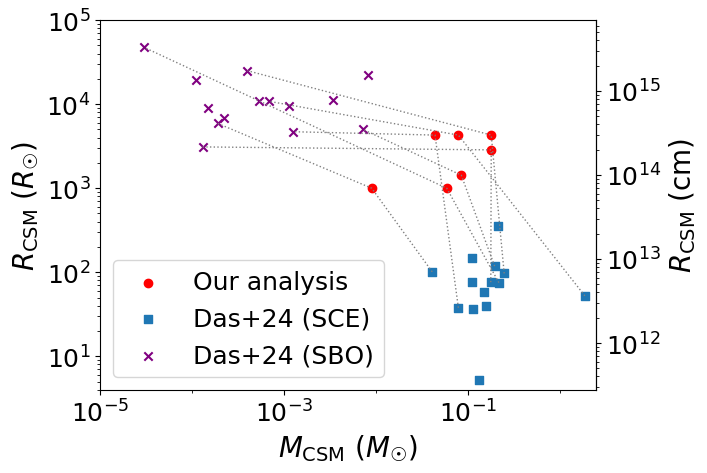}
    \caption{
    CSM parameters obtained in this work by fitting to hydrodynamic modelling for the SNe in Table~\ref{tab:sample_info} (red), along with CSM parameters estimated by \citet{Das2024-pm} for their set of SNe, using either order-of-magnitude estimates of shock breakout emission (purple) or the analytical model of shock cooling emission from extended envelope by \citet{Piro2021-cu} (blue).
    Data points corresponding to the same SN are respectively connected with grey dotted lines.
    Blue data points with grey labels correspond to SNe which are not included in our sample.}
    \label{fig:CSM_params}
\end{figure}


Figure~\ref{fig:lc_all} shows each of the best fit light curves.
Table~\ref{tab:sample_info} presents the estimated CSM properties and Figure~\ref{fig:CSM_params} shows $R_\CSM$ and $M_\CSM$ that yield the best fit between the synthetic and observed light curves.
The ranges of $M_\CSM$ and $R_\CSM$ derived in this study are $M_\CSM = 0.01 - 0.2 \ \Msun$ and $R_\CSM = (0.7 - 3) \times 10^{14} \ \mathrm{cm}$, respectively.

Figure~\ref{fig:CSM_params} also shows the parameters obtained in the previous analysis by \citet{Das2024-pm}.
They considered two possibilities where either shock breakout or shock cooling is responsible for the early peaks.
For the case of shock breakout, they derived order-of-magnitude estimates, while for the case of shock cooling, the CSM configurations were estimated with light curve fitting to the analytical model of \citet{Piro2021-cu}.
Figure~\ref{fig:lc_all} shows that shock breakout and cooling emission are both present in most of our models, but the observational data points of early peaks are mostly fit with the contribution from shock cooling emission.
Comparing between the CSM configurations inferred here and those derived in \citet{Das2024-pm} with the assumption of shock cooling emission, we see that the estimated values of $M_\CSM$ (largely corresponding to peak timescale) are similar, while $R_\CSM$ differs by a factor of $\sim 10$.
This can be attributed to the simplifying assumption in blackbody temperature used in analytical models, as noted in Section~\ref{sec:dependence}.

A similar trend can also be seen in comparison with previous results which similarly fit the early peaks of objects in our sample using analytical models of shock cooling emission from an extended envelope.
\citet{Ho2023-qn} reported the limits of $M_e < 10^{-2} \ \Msun$, $R_e > 10^{12} \ \mathrm{cm}$ for SN 2020bvc.
The parameters were estimated not through a light curve fit, but through analytical estimates of peak timescale and luminosity \citep{Nakar2014-yn}.
Since the light curve only contained the declining part of the early peak, they did not resolve the degeneracy between the mass and radius of the shocked material.
\citet{Jacobson-Galan2022-db} obtained the values of $M_e = (0.013 - 0.47) \times 10^{-2} \ \Msun$ and $R_e = 27.5 - 385 \ \Rsun$ for SN 2021gno.
These results were obtained by fitting the light curve to various theoretical models of shock cooling emission from an extended envelope \citep{Piro2015-nm, Sapir2017-pw, Piro2021-cu}.

\begin{figure}
    \centering
    \includegraphics[width=1\linewidth]{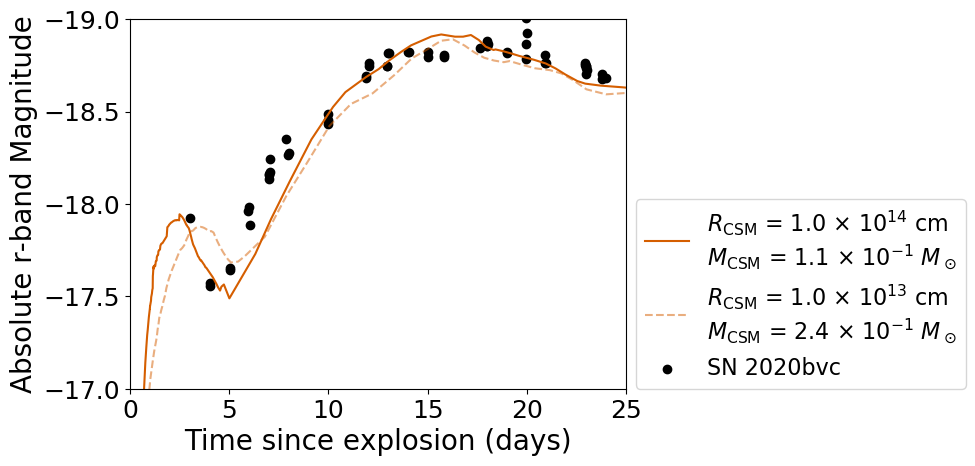}
    \caption{
    $r$-band light curve of SN 2020bvc and its theoretical models for $R_\CSM = 10^{14} \ \mathrm{cm}$ (solid) and $R_\CSM = 10^{13} \ \mathrm{cm}$ (dashed).
    The model with $R_\CSM = 10^{13} \ \mathrm{cm}$ results in an early peak on a longer timescale ($\sim 4 \ \mathrm{d}$) than the observed light curve ($< 3 \ \mathrm{d}$).
    }
    \label{fig:test_degeneracy}
\end{figure}

Our results are in broad agreement with those of \citet{Jin2021-hq}, who modelled the early peaks observed in three Type Ic SNe (including SN 2020bvc, which is also in our sample) with CSM interaction using \texttt{STELLA}.
However, they also report two similarly acceptable fits for SN 2020bvc: $M_\CSM = 0.1 \ \Msun$, $R_\CSM = 10^{14} \ \mathrm{cm}$ and $M_\CSM = 0.2 \ \Msun$, $R_\CSM = 10^{13} \ \mathrm{cm}$.
To assess the degree of this degeneracy between $M_\CSM$ and $R_\CSM$, we constructed additional models of SN 2020bvc with finer gridding of $\dot{M}$ for $R_\CSM = 10^{13} \ \mathrm{cm}$ and $R_\CSM = 10^{14} \ \mathrm{cm}$.
Figure~\ref{fig:test_degeneracy} shows the best-fit light curves among the additional models.
While they exhibit early peaks with similar luminosities, we find that the light curves corresponding to such widely different values of $R_\CSM$ are distinguishable given sufficiently well-sampled light curve data.
In particular, the model with $R_\CSM = 10^{13} \ \mathrm{cm}$ exhibits an early peak on a longer timescale ($\sim 4 \ \mathrm{d}$) than the observed light curve ($< 3 \ \mathrm{d}$). 
Therefore, we consider the degeneracy between $M_\CSM$ and $R_\CSM$ to be less severe than previously reported.

Several sources of uncertainty affect our parameter estimates.
While uncertainties exist in the CSM density profile, it has shown that the change in the CSM density profile does not affect the light curve relative as much as CSM masses and radii \citep{Jin2021-hq}.
The largest source of uncertainty likely stems from our choice of $E_\ej$.
The factor of $\sim 1.5$ difference in $M_\ej$ between analytical estimate and best-fit values, discussed in Section~\ref{sec:analysis}, provides a rough estimate of the uncertainty in this parameter, which propagates to the inferred CSM parameters with similar relative magnitude.
Combined with the coarse sampling within the parameter space of CSM configuration (adjacent values $\dot{M}$ and $R_\CSM$ in the grid differ by factors of $1.5 - 2$), we consider our estimates to have an uncertainty up to a factor of a few to the inferred parameters.

\section{Discussion}
\label{sec:discussion}

\subsection{Scenarios for CSM formation}

Based on the parameters we derived in this study, we discuss various physical mechanisms that could lead to the formation of CSM responsible for the early peaks.

The inferred values of $\dot{M}$ are too high to be explained by stellar winds, even for stars with high wind mass loss rates like Wolf--Rayet stars \citep[e.g.,][]{Smith2014-ci}. 
Recent numerical studies on wave-driven mass loss predict the CSM masses of at most $10^{-2} \ \Msun$ \citep{Fuller2018-ya, Leung2021-dp}, which are also insufficient to account for large values of $M_\CSM$ we obtained.

The violent runaway of off-centre silicon burning is also proposed as a channel for CSM formation \citep{Woosley2019-ke}.
This degenerate silicon flash is expected to only occur for relatively light He stars (initial mass of $2.5 - 3.2 \ \Msun$).
While the predicted masses of ejected material that can turn into CSM ($0.02 - 0.74 \ \Msun$) align with our results, this scenario currently lacks predictions for CSM radii.

The stripping of H-rich envelopes from the progenitors of Type Ibc SNe is expected to occur through Case B mass transfer to their binary companions \citep[e.g.,][]{Podsiadlowski1992-px}.
Several later mass transfer episodes can give rise to CSM around the progenitors of Type Ibc SNe. 
Case BB mass transfer is expected to generate only a moderate amount of CSM, which does not reach the values reported in Table~\ref{tab:sample_info} \citep[e.g.,][]{Laplace2020-il}. 
However, even later episodes of mass transfer, such as Case BC mass transfer, and ``Case X'' mass transfer (mass transfer after core silicon depletion, a term coined by \citealt{Ercolino2025-zx}), could account for the CSM masses of up to $0.2 \ \Msun$ that we obtained \citep[see also][]{Wu2022-ek}. 
The values of $R_\CSM$ reported in Table~\ref{tab:sample_info} are roughly consistent with the estimate of the circumbinary disc (CBD) radii by \citet{Ercolino2025-zx} \citep[see also][]{Tuna2023-kh}, if the binary is tight enough (period $\lesssim \qty{10}{d}$).
A further consideration of this CBD scenario requires a better understanding of the highly uncertain density profile and geometry of the CBD, as well as their effect on the early peaks, which we leave to future studies.
In addition, it should be noted that these late-time mass transfer episodes only occur in stars with light helium-rich envelopes; for example, \citet{Ercolino2025-zx} found that the largest initial mass of helium star that can undergo Case BC mass transfer is $4.73 \ \Msun$.
Therefore, while this scenario is consistent with most of $M_\ej$ in Table~\ref{tab:sample_info}, it is at odds with the value of $M_\ej = 7 \ \Msun$ reported for SN 2021nng, and an alternative scenario should be considered.

%

Very massive stars with helium core masses of $30 - 60 \ \Msun$ are predicted to experience mass loss due to stellar pulses caused by pair instability in their cores \citep[see e.g.,][]{Woosley2007-ti}.
Hydrodynamic modelling predicts that pair-instability pulses can generate CSM with masses ranging from $\sim 10^{-2} \ \Msun$ to more than $\sim 10 \ \Msun$ \citep{Woosley2017-bo, Leung2019-rk, Renzo2020-sq}.
\citet{Renzo2020-sq} summarise the ejected mass, the velocity of its centre of mass, and the duration until the core-collapse for the pulses observed in their models.
Since the velocity of the material ejected by pulses is estimated to be $1000 - \qty{3000}{km.s^{-1}}$, the values of $R_\CSM$ reported in  Table~\ref{tab:sample_info} correspond to mass ejection occurring $\sim 10^{-2} \si{yr}$ before the core collapse.
This time duration and the values of $M_\CSM$ reported in  Table~\ref{tab:sample_info} correspond to models with an initial helium core mass within a rather narrow range of $52.25 - 53.25 \ \Msun$ in \citet{Renzo2020-sq}.
The plausibility of such an explosion occurring, its observability, and its resemblance to a normal Type Ib SN are all dubious.
In addition to the stringent restriction on mass as well as metallicity, there is an additional difficulty as the model predicts that the stars in the aforementioned mass range will directly collapse into a black hole; PPISNe only ocurred for initial helium core masses larger than $\sim 80 \ \Msun$.
The estimated ejecta mass $M_\ej = 7 \ \Msun$ could be reconciled with very massive progenitors if only a small fraction of material is ejected and the rest falls back onto the black hole, as seen in the simulation of \citet{Chan2018-up}.
Further detections of early peaks in Type Ibc SNe with relatively high ejecta mass will enable us to elaborate on this hypothesis.

\subsection{Shock cooling emission from CSM and extended envelope}

In this work, we did not consider the scenario in which an extended stellar envelope is responsible for the early peaks.
Robustly distinguishing between shock cooling emission from CSM and that from an extended envelope is challenging, since they arise from identical physical processes.
Recently, \citet{Haynie2025-gs} demonstrated that shock cooling light curve can exhibit two distinct phases corresponding to emission from CSM and extended envelope, respectively. 
They found that since the envelope is heavier, the timescale of shock cooling emission from the envelope is longer than that from CSM, making it able to distinguish the two components.
The authors also showed that the first peak in the double-peaked bolometric light curve of ultra-stripped SN 2019dge can be well reproduced numerically by an explosion of a model with an extended helium envelope and CSM created through expansion and mass transfer after the star’s oxygen/neon burning phase and onwards \citep{Wu2022-ek}.
Further detections of SNe with similar light curve features will lead to better characterisation of the two processes.
They also calls for the proper treatment of the enhancement to $\radni$ heating peaks due to shock cooling emission; to that end, and in order to adequately account for the diverse density profiles of CSM, the model grid that ensures the consistent realisation of both components of shock cooling as well as radioactive heating peaks will be useful, which we leave to future works.

\subsection{Prospects of detection by future UV surveys}

\begin{figure}
    \centering
    \includegraphics[width=1\linewidth]{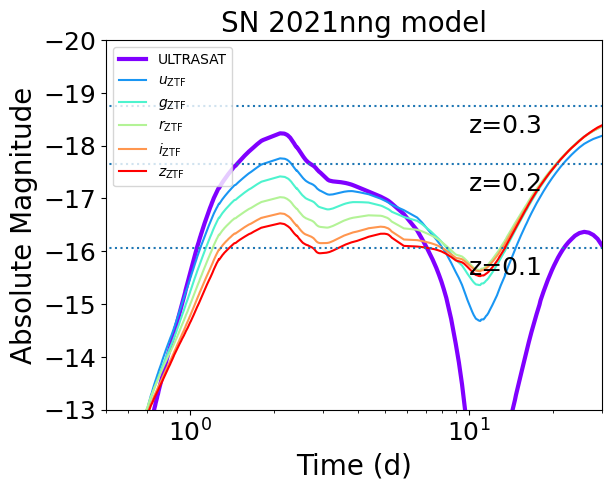}
    \caption{Model light curves of a shock cooling peak from a Type Ibc SN seen through \textit{ULTRASAT} filter (purple) and ZTF filters, calculated for our best-fit model of SN 2021nng. 
    Dotted lines indicate $5 \sigma$ detection limit of \textit{ULTRASAT} at 900 seconds integration \citep[$22.5 \ \mathrm{mag}$,][]{Shvartzvald2024-un} for redshifts $z = 0.1, 0.2, 0.3$.}
    \label{fig:ultrasat}
\end{figure}

Since emission powered by interaction with dense CSM or shock cooling emission from an extended envelope is typically bright in the ultraviolet (UV) wavelength range, it is an interesting target for a high-cadence UV survey, such as the planned survey by \textit{ULTRASAT} \citep{Shvartzvald2024-un}. 
Figure~\ref{fig:ultrasat} shows an example of simulated \textit{ULTRASAT} light curve using \textit{ULTRASAT}'s total throughput from \citet{Shvartzvald2024-un}. 
The figure also shows the absolute magnitudes corresponding to the nominal detection limit of \textit{ULTRASAT} (22.5 mag) at redshifts $z = 0.1, 0.2, 0.3$.
The early peaks will be observable by \textit{ULTRASAT} up to around $z \sim 0.2$. 
Since the \textit{ULTRASAT} high-cadence survey will observe an area of $\qty{200}{deg^2}$, the volumetric rate of Type Ibc SNe in the local Universe is estimated at $\qty{2.5e-5}{Mpc^{-3}.yr^{-1}}$ \citep{Li2011-jx}, and the fraction of Type Ibc SNe with early peaks is estimated by \citet{Das2024-pm} as $\sim 5 \%$, we estimate that during three years of the \textit{ULTRASAT} high-cadence survey, $\sim 30$ Type Ibc SNe with early peaks will occur within $z < 0.2$ and within \textit{ULTRASAT}'s field of view.


\section{Conclusion}
\label{sec:conclusion}

In this study, we performed hydrodynamic simulations of Type Ibc SNe interacting with confined, dense CSM. 
We derived the radii and masses of the CSM, which ranged from $(1 - 5) \times 10^3 \ \Rsun$ and $10^{-2} - 10^{-1} \ \Msun$, respectively.
There is a notable discrepancy in CSM radii compared to the results of the previous analyses, which is attributed to the simplifying assumption made regarding to the blackbody temperature of the emission in analytic models.
We also inferred that, if confined CSM is responsible for the early peaks, CSM might have been created either as a CBD formed by late-time mass transfer episodes, or mass ejection due to pulsational pair instability, although the latter possibility has some difficulty concerning its plausibility and detectability.

The relationship between different CSM density profiles and the resulting shock cooling light curves, which is known not to manifest itself in the light curves, remains unclear.
Further numerical modelling of CSM interaction is required to better account for the expected diversity of CSM density profiles as well as the asphericity of CSM; this will enable us to investigate the mechanisms of CSM formation in greater detail.
With the advent of future high-cadence UV surveys, such as \textit{ULTRASAT}, providing prime opportunities to observe early peaks in SN light curves in larger numbers, it is crucial to refine our understanding of SNe interacting with confined CSM through detailed numerical characterization of these signatures.

\section*{Acknowledgements}

We thank Joe Anderson, R{\'e}gis Cartier, Francisco F{\"o}rster, Giuliano Pignata, Jos{\'e} Prieto, Christopher Irwin, Tomoya Kinugawa, Ryo Sawada, and Dan Kasen for valuable comments regarding this work.
This work is supported by the Grants-in-Aid for Scientific Research of the Japan Society for the Promotion of Science (JP24K00682, JP24H01824, JP21H04997, JP24H00002, JP24H00027, JP24K00668) and by the Australian Research Council (ARC) through the ARC's Discovery Projects funding scheme (project DP240101786).
Numerical computations were in part carried out on the PC cluster at the Yukawa Institute for Theoretical Physics (YITP), Kyoto University, and at the Center for Computational Astrophysics (CfCA), National Astronomical Observatory of Japan.

\section*{Data Availability}

The data underlying this article will be shared on reasonable request to the corresponding author.



\bibliographystyle{mnras}
\bibliography{paperpile}



\appendix

\section{Verification of the simplifying assumptions}
\label{sec:Jin21_test}

\begin{figure}
    \centering
    \includegraphics[width=1\linewidth]{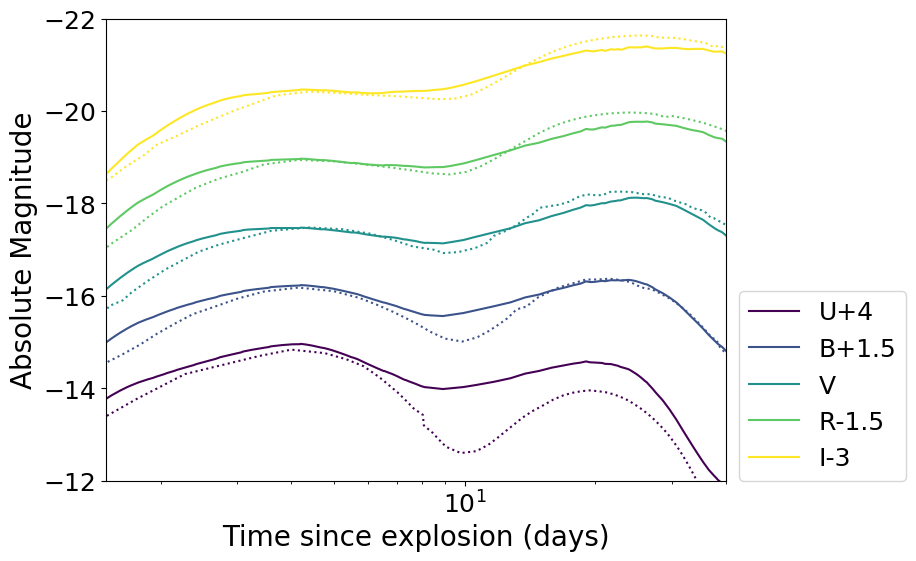}
    \caption{The result of our test calculation \texttt{4P\_fm0.15\_E2\_0.15M\_R14} model of \citet{Jin2021-hq}, along with their results using thermal bomb method.}
    \label{fig:Jin21_test}
\end{figure}

Figure~\ref{fig:Jin21_test} shows results from our test calculation with the parameters chosen to match the \texttt{4P\_fm0.15\_E2\_0.15M\_R14} model from \citet{Jin2021-hq}, who exploded the progenitor models with the thermal bomb method.
Along with the result of our calculation, we also show light curves taken from Figure~3 of \citet{Jin2021-hq}.
The figure shows that our simplified model can calculate the early, interaction-powered peaks in a manner that is mostly consistent with a more detailed calculation.
We can also see that the radioactively powered peaks are also consistent in the visual bands, whose wavelengths are similar to those we mainly used to fit the models without CSM.
The discrepancy in the $U$-band, which likely arises from the difference in ejecta density structure, is not significant to the discussion presented in this paper.
We therefore consider our model to be sufficiently consistent with more sophisticated models. 

\bsp	
\label{lastpage}
\end{document}